\documentclass[authoryear]{elsarticle}
\usepackage[a4paper, total={6in, 10in}]{geometry}
\usepackage{booktabs}
\usepackage{float}
\usepackage{amsmath}
\usepackage{amsthm}
\usepackage{amsmath,amssymb,amsfonts,mathrsfs,hyperref,microtype, amsthm}
\usepackage{graphicx}
\usepackage{graphicx,mathabx}
\usepackage{enumitem}  
\graphicspath{ {images/} }
\numberwithin{equation}{section}

\newtheorem*{example*}{Example}

\newcommand{\Rmnum}[1]{\expandafter\@slowromancap\romannumeral #1@}

\journal{}
\makeatletter
\def\ps@pprintTitle{%
   \let\@oddhead\@empty
   \let\@evenhead\@empty
   \def\@oddfoot{\reset@font\hfil\thepage\hfil}
   \let\@evenfoot\@oddfoot
}

\begin{document}

\begin{frontmatter}
\author[a]{Helmi Shat\corref{cor1}}
\ead{hshat@ovgu.de}
\author[a]{Rainer Schwabe}
\ead{rainer.schwabe@ovgu.de}
\address[a]{\small Institute for Mathematical Stochastics, Otto-von-Guericke University Magdeburg, \\ Universit\"atsplatz 2, 39106 Magdeburg, Germany }

 \title{Optimal Stress Levels in Accelerated Degradation Testing for Various Degradation Models}

\begin{abstract}
Accelerated degradation tests are used to provide accurate estimation of lifetime characteristics of highly reliable products within a relatively short testing time. Data from particular tests at high levels of stress (e.g., temperature, voltage, or vibration) are extrapolated, through a physically meaningful statistical model, to attain estimates of lifetime quantiles at normal use conditions.
The gamma process is a natural model for estimating the degradation increments over certain degradation paths, which exhibit a monotone and strictly increasing degradation pattern.
In this work, we derive first an algorithm-based optimal design for a repeated measures degradation test with single failure mode that corresponds to a single response component. The univariate degradation process is expressed using a gamma model where a generalized linear model  is introduced to facilitate the derivation of an optimal design. Consequently, we extend the univariate model and characterize optimal designs for accelerated degradation tests with bivariate degradation processes. The first bivariate model includes two gamma processes as marginal degradation models. The second bivariate models is expressed by a gamma process along with a mixed effects linear model.
We derive optimal designs for minimizing the asymptotic variance for estimating some quantile of the failure time distribution at the normal use conditions. Sensitivity analysis is conducted to study the behavior of the resulting optimal designs under misspecifications of adopted nominal values.
\end{abstract}

\begin{keyword}
Accelerated degradation test\sep gamma model\sep linear mixed-effects model\sep the multiplicative algorithm \sep locally c-optimal design.

\end{keyword}

\end{frontmatter}

\section{Introduction}
\label{sec:1}
Along with the huge advances of industrial technologies, the companies are forced to manufacture highly reliable products in order to compete in the industrial market. During the design stage, it is extremely significant to assess the reliability related properties of the product. One of the proposed methods to handle this issue is accelerated life testing (ALT). However, it
is difficult to obtain enough failure time data to satisfy the
requirement of ALT because of the high-reliable property of
products. Hence,  accelerated degradation test (ADT) is
suggested in order to give estimations in relatively short periods of time about the life time and reliability of the system under study. ADT might be divided into three classes, constant stress ADT (CSADT), step
stress ADT (SSADT) and progressive ADT. In our model, we consider the optimal planning of CSADT where the testing units are divided into groups where each group is tested under distinct stress combination.
Numerous researches have considered the implementation of ADT to provide reliability estimations. \cite{7114339} derived an algorithm-based optimal
ADT procedure by minimizing the asymptotic variance
of the MLE of the mean time to failure of a product, where
the sample size and termination time of each run of the ADT at a
constant measurement frequency were determined. \cite{ZHANG2015369} suggested an
analytical optimal ADT design method for more efficient reliability demonstration by minimizing the
asymptotic variance of decision variable in reliability demonstration under the constraints of sample size, test duration, test cost, and predetermined decision risks. Considering linear mixed effects model (LMEM), \citep{doi:10.1002/asmb.2061} utilized also the minimum asymptotic variance criterion to develop optimal design as well as compromise design plans for accelerated degradation tests. Further, \citep{10.2307/24307139} provided $D$-optimal experimental designs for the estimation of fixed effects and two variance components, in the presence of nested random effects. The authors show that the designs when the samples are distributed as uniformly as possible among batches result in $D$-optimal designs for maximum likelihood estimation. For the non-linear case, \citep{doi:10.1111/biom.12660} presented $D$-optimal experimental designs for non-linear mixed effects models, where a categorical factor with covariate information is a design variable combined with another design factor. Moreover, \citep{SINHA20111394} studied the performance of the locally $D$-optimal sequential designs for analyzing generalized linear mixed models. The authors demonstrate that one could attain considerable gain in efficiency from the maximum likelihood estimators when data are augmented with the  sequential design scheme rather than the much simpler uniform design scheme. 
Considering Gamma process models, \citep{tsai2012optimal} discussed the problem
of cost-constrained optimal design for degradation tests based on a gamma degradation process with random effects. The authors provide further an analytical assessment of the effects of
model mis-specification that occur when the random effects are
not taken into consideration in the gamma degradation model. 
 In addition, \citep{doi:10.1080/03610918.2012.700749} introduced reliability model of the degradation products with two performance characteristics based on a Gamma process, and then present the
corresponding SSADT model. Next, under the constraint of total experimental cost, the optimal settings such as sample size, measurement times, and measurement frequency are obtained by minimizing the asymptotic variance of the estimated 100$\alpha$th percentile of the product’s lifetime distribution.
In order to predict the lifetime of the population from ADT, \citep{WANG2015172} considered Gamma process with a time transformation and random effects. They present a deducing method for determining the relationships between the shape and scale parameters of Gamma process and accelerated stresses. \cite{doi:10.1080/03610926.2018.1459718} discussed optimal design problems for CSADT based on Gamma processes with
fixed effect and random effect. They prove
that, for $D$-optimality,$V$-optimality and $A$-optimality criteria, optimal CSADT plans with multiple stress levels degenerate to two stress-level test plans only using the minimum and maximum stress levels under model assumptions. \cite{e17052556} developed statistical methods for optimal designing ADT plans under the total experimental cost constraint and assuming that the degradation characteristic follows a Gamma process model. In addition, the author  derives compromise plans to provide means to check the adequacy of the assumed
acceleration model. ADT with the presence of Competing failure modes is an important reliability area to be addressed. Therefore, the study of the statistical inference of ADT with competing failures is of great significance. 
\citep{7160792} introduced a modeling approach to simultaneously analyze linear degradation data and traumatic failures with competing risks in an SSADT experiment. Moreover, methodology for ALT planning when there are two or more independent failure modes was discussed by \citep{4118444}. The author assumed that the failure modes have respective latent failure times, and the minimum of these times corresponds to the product lifetime. The latent failure times are assumed to be
-independently distributed Weibull with known, common shape
parameter. Considering accelerated destructive degradation tests (ADDT), \citep{doi:10.1002/9781118826805.ch22} proposed methods to find unconstrained and constrained optimum test plans for competing risk applications under a $V$-optimality criterion that aim to minimize the large-sample approximate variance of a failure-time distribution quantile at use conditions. The authors consider linearly degraded response models with an application for an adhesive bond. In regards to nonparameteric methods of evaluation, 
\cite{balakrishnan2019nonparametric} introduced some approximation techniques of the first passage time distribution of the degradation processes incorporating random effects if the process type is unknown. The authors approximate the density function of some stochastic degradation processes, i.\,e. Gamma process and inverse Gaussian process, by inverting the empirical Laplace transform using the empirical saddlepoint method. \cite{palayangoda2020improved} extended the work of \citep{balakrishnan2019nonparametric} and proposed some improved
techniques based on saddlepoint approximation where numerical examples and Monte Carlo simulation studies are used to illustrate the advantages of the proposed techniques. Further, 
\cite{balakrishnan2017gamma} considered the theoretical aspects as
well as the application of Gamma processes in degradation analysis. The authors give some statistical properties of degradation models based on Gamma processes under different tests.\\

The rest of this article is organized as follows. Section~\ref{sec:11} is devoted to develop optimal experimental designs for a univariate gamma model.  In section \ref{sec:222} we introduce an optimal design considering a bivariate gamma process where the corresponding failure modes do not interact. In Section~\ref{sec:2}, we characterize a $c$-optimal design for an ADT with a bivariate degradation model with repeated measures given that one of the marginal response components follows a gamma process model where the other follows a LMEM. The paper closes with a short discussion in Section~\ref{sec1.4444}. All numerical computations were made by using the R programming language\citep{R}.

\section{Accelerated degradation testing based on a gamma process}
\label{sec:11}
The gamma process is a natural stochastic model for degradation processes in which degradation is assumed to occur gradually over time in a sequence of independent increments. 
In this section, we assume that the testing unit has a single dependent failure mode where the degradation path is characterized by a gamma process model in terms of a standardized time variable $t$. 
In addition, it is assumed that there is a single stress variable and its (standardized) stress level $x$ can be chosen by the experimenter from the experimental  region $\mathcal{{X}} = [0,1]$. 
The subsequent subsections clarify the approximation of the gamma model with a generalized linear model approach. 
Further, we explain the derivation of the corresponding information matrix in order to obtain an algorithm based optimal experimental design with respect to the asymptotic variance of a quantile of the failure time distribution. 

\subsection{Model formulation}
\label{sec:1.2}
A gamma process $Z_t$ is a stochastic process with independent gamma distributed increments. 
The process can be parameterized by the rate $\gamma$ and a scale parameter $\nu$. If the process is observed at $k$ subsequent time points $t_j$, $0 < t_1 < ... < t_k$, then the $j$th degradation increment $Y_{j} =  Z_{t_j} - Z_{t_{j - 1}}$ is gamma distributed with shape $\gamma \Delta_{j}$ and scale $\nu$, where $\Delta_j = t_j - t_{j - 1}$ is the length of the $j$th time interval and $Z_{t_0} = 0$ at $t_0 = 0$.

We assume that the stress variable $x$ only affects the rate $\gamma = \gamma(x)$ of the gamma process and, hence, the shape parameters $\gamma(x) \Delta_{j}$ of the increments while the scale parameter $\nu$ is fixed and known. 
We further assume that the rate $\gamma(x)$ is given by a linear trend in the stress variable with a logarithmic link, 
\begin{equation} 
\label{shape}
\gamma(x) = e^{\beta_{0} + \beta_{1} x} ,
\end{equation}
where the intercept $\beta_0$ and the slope $\beta_1$ are to be estimated.
When a unit is tested under stress level $x$ during a time interval of length $\Delta$, the degradation increment $Y$ has density
\begin{equation}
	\label{marginal}
	f_Y(y) = \frac{y^{\gamma(x) \Delta - 1} e^{-y / \nu}}{\Gamma(\gamma(x) \Delta) \nu^{\gamma(x) \Delta}},
\end{equation}
where $\Gamma(\alpha) = \int_0^\infty z^{\alpha - 1} e^{ - z} \mathrm{d}z$ is the gamma function.
The mean of the increment is given by
\begin{equation} 
	\label{expected-value-gamma}
	\mu(x) = \mathrm{E}(Y) = \gamma(x) \Delta \nu = e^{\beta_{0} + \beta_{1} x} \Delta \nu .
\end{equation}
Thus the mean $\mu(x)$ is linked to the linear predictor $\beta_0 + \beta_1 x$ by a scaled log link.
Hence, the model assumptions fit into the concept of generalized linear models.

To be more specific, in accelerated degradation testing $n$ distinct testing units are tested at potentially different stress settings $x_i$ which are held fixed over time for each unit $i = 1, ..., n$.
Measurements are made at  predetermined time points $t_1, ..., t_k$ which are identical for all units. 
The degradation increments $Y_{ij}$ when testing unit $i$ during the $j$th time interval of length $\Delta_j = t_j - t_{j - 1}$ are independent gamma distributed with shape $\gamma(x_{i}) \Delta_{j}$ and scale $\nu$.

\subsection{Estimation and information}
\label{sec:1.3}
Denote by $\boldsymbol{\beta} = (\beta_0, \beta_1)^T$ the vector of unknown parameters. 
By \eqref{marginal} the log-likelihood of a single degradation increment $Y$ is given by
 \begin{equation}
		\label{eq-loglik-single}
		\ell(\boldsymbol{\beta}; y) = (e^{\beta_{0} + \beta_{1} x} \Delta - 1) \ln(y)  - y / \nu - \ln\left(\Gamma(e^{\beta_{0} + \beta_{1} x} \Delta)\right) - e^{\beta_{0} + \beta_{1} x} \Delta \ln(\nu) 
\end{equation}
when the stress level $x$ is applied and the increment is measured over a time interval of length $\Delta$.
The elemental Fisher information matrix $\mathbf{M}_{\boldsymbol\beta}(x, \Delta)$ related to a single increment can be calculated as minus the matrix of expected second order derivatives of the log-likelihood, 
\begin{equation}
	\label{eq-info-elemental-single}
	\mathbf{M}_{\boldsymbol\beta}(x, \Delta)
	= q(\beta_{0} + \beta_{1} x + \ln(\Delta)) \left(
	\begin{array}{cc}
		1 & x
		\\
		x & x^2
	\end{array}
	\right) ,
\end{equation}
where $q$ is defined by $q(z) = e^{2z} \psi_1(e^z)$ and $\psi_1(\alpha) = \mathrm{d}^2 \ln\left(\Gamma(\alpha)\right) / \mathrm{d}\alpha^2$ is the trigamma function.

Because the increments $Y_{i1}, ..., Y_{ik}$ measured at times $t_1,...,t_k$ are statistically independent within a unit $i$, the log-likelihood $\ell(\boldsymbol{\beta}; y_{i1}, ..., y_{ik}) = \sum_{j = 1}^{k} \ell(\boldsymbol{\beta}; y_{ij})$ of a unit $i$ is the sum of the log-likelihoods for the single observations $Y_{ij}$.
Thus also the information matrix $\mathbf{M}_{\boldsymbol\beta}(x_i)$ of a unit is the sum of the information of the single increments, 
\begin{equation}
	\label{eq-info-elemental-unit}
	\mathbf{M}_{\boldsymbol\beta}(x_i)
	= \sum_{j = 1}^{k} \mathbf{M}_{\boldsymbol\beta}(x_i, \Delta_j)
	= \lambda(\beta_{0} + \beta_{1} x_i) \left(
	\begin{array}{cc}
		1 & x_i
		\\
		x_i & x_i^2
	\end{array}
	\right) ,
\end{equation}
where the ``intensity'' $\lambda(z) = \sum_{j = 1}^{k} q\big(z + \ln(\Delta_j)\big)$ accounts for the contribution of the non-linearity at $z = \beta_0 + \beta_1 x_i$ to the information.

Furthermore, because measurements are statistically independent between units, both the log-likelihood $\ell(\boldsymbol{\beta}; y_{11}, ..., y_{nk}) = \sum_{i = 1}^{n} \ell(\boldsymbol{\beta}; y_{i1}, ..., y_{ik})$ and the information
\begin{equation}
	\label{eq-info-exact}
	\mathbf{M}_{\boldsymbol\beta}(x_1, ..., x_n)
	= \sum_{i=1}^{n} \mathbf{M}_{\boldsymbol\beta}(x_i)
\end{equation}
for the whole experiment summarize the log-likelihood and the information of the units.
This information matrix $\mathbf{M}_{\boldsymbol\beta}(x_1,...,x_n)$ provides a measure for the performance of the experiment as its inverse is proportional to the asymptotic variance covariance matrix for the maximum likelihood estimator of $\boldsymbol{\beta}$.

In an accelerated degradation experiment the stress variable $x$ is under control of the experimenter.
For each unit $i$, the setting $x_i$ of the stress variable adjusted to $i$ may be chosen from an experimental region $\mathcal{X}$.
The collection $x_1, ..., x_n$ of these settings is called the design of the experiment.
An optimal design then aims at minimizing an optimality criterion which is a function of the information matrix.

Finding optimal designs $x_1, ..., x_n$ is, in general, a difficult task of discrete optimization. 
To circumvent this problem we follow the approach of approximate designs propagated by \cite{kiefer1959optimum}. 
For this first note that by \eqref{eq-info-exact} the information matrix $\mathbf{M}_{\boldsymbol\beta}(x_1,...,x_n) = \sum_{i=1}^{m} n_i \mathbf{M}_{\boldsymbol\beta}(x_i)$ does only depend on the set of mutually distinct settings $x_1,...,x_m$, say, in the design and their corresponding frequencies $n_1,...,n_m$, $\sum_{i=1}^m n_i = n$.
For approximate designs the requirement of integer numbers $n_i$ of testing units at stress level $x_i$ is relaxed. Then methods of continuous convex optimization can be employed to find optimal designs, see \citep{silvey1980optimal}, and efficient designs with integer numbers $n_i$ can be derived by proper rounding the optimal solutions to nearest integers. 
For this approach the sample size $n$ does not play a role on the optimization step when proportions $w_i = n_i/n$ are considered.
This approach is, in particular, of use when the number $n$ of units is sufficiently large which is appropriate in the present non-linear setup, where asymptotic performance is measured. 
Moreover, in this approach, the frequencies $n_i$ will be replaced by proportions $w_i=n_i/n$, because the total number $n$ of units does not play a role in the optimization.
Thus an approximate design $\xi$ is defined by a finite number of settings $x_i$ from the experimental region $\mathcal{{X}}$ with associated weights $w_i > 0$, $i = 1,...,m$, $\sum_{i=1}^m w_i = 1$. 
Accordingly, the corresponding standardized, per unit information matrix is defined as 
\begin{equation}
	\label{eq-info-approximate}
	\mathbf{M}_{\boldsymbol\beta}(\xi) = \sum_{i=1}^{m} w_i \mathbf{M}_{\boldsymbol\beta}(x_i)
\end{equation}
so that ``exact'' designs $x_1, ..., x_n$ are properly embedded by $\mathbf{M}_{\boldsymbol\beta}(\xi) = (1 / n) \mathbf{M}_{\boldsymbol\beta}(x_1, ..., x_n)$.

As the information matrix depends on the parameter vector $\boldsymbol{\beta}$ only through the linear predictor $\beta_0 + \beta_1 x$, a canonical transformation can be employed which simultaneously maps experimental settings $x$ to $z = \beta_0 + \beta_1 x$ and the parameters $\beta_0$ and $\beta_1$ to the standardized value $\beta_0 = 0$ and $\beta_1 = 1$ for analytical solutions, see \cite{ford1992use}.

When all time intervals have the same length $\Delta_j = \Delta$, $j = 1, ..., k$, the influence of the repeated measurements reduces to $\lambda(z) = k \, q\big(z + \ln(\Delta)\big)$ for the intensity and, hence, to a multiplicative factor $k$ in the information matrix. Thus, for common design criteria, the number $k$ of measurements is immaterial for design optimization.

\subsection{Optimality criterion based on the failure time distribution}
\label{sec:1.345}
In degradation testing we are interested in characteristics of the failure time distribution of soft failure due to degradation under normal use condition ${x}_u$.
It is supposed that the gamma process $Z_{u,t}$ describing the degradation under normal use condition has the rate $\gamma(x_u) = \exp(\beta_0 + \beta_1 x_u)$ as in equation (\ref{shape}) and scale $\nu$. 
Typically the normal use condition $x_{u}$ is not contained in the experimental region $\mathcal{X}$, $x_{u} < 0$. 
Further it is natural to assume that the degradation paths are strictly increasing over time. 
Then a soft failure due to degradation is defined as exceedance of the degradation path over a failure threshold $z_{0}$. 
The failure time $T$ under normal use condition is defined as the first time $t$ the degradation path $Z_{u,t}$ reaches or exceeds the threshold $z_{0}$, i.\,e., $T = \inf \{t \geq 0;\, Z_{u,t} \geq z_{0}\}$. 
In order to derive certain characteristics of the distribution of the failure time, we determine its distribution function $F_T(t) = \mathrm{P}(T \leq t)$. 
For this note that $T \leq t$ if and only if $Z_{u,t} \geq z_{0}$. 
The degradation $Z_{u,t}$ at time $t$ is gamma distributed with shape $\gamma(x_{u}) t$ and scale $\nu)$.
Hence, the distribution function of the failure time $T$ can be expressed as
\begin{equation}
\label{eq-failure-time-distribution}
\begin{split}
F_{T}(t) & = \mathrm{P}(Z_{u,t} \geq z_{0}) \\
	& = \frac{1}{\Gamma(\gamma(x_{u}) t)} \int_{z_{0}}^{\infty} (z / \nu)^{\gamma(x_{u}) t - 1} e^{ - z / \nu} \nu^{ - 1} \mathrm{d}z 
	\\
	& = Q(\gamma(x_{u}) t , z_{0} / \nu)
\end{split}
\end{equation}
where $Q(s, z) = \Gamma(s, z)/\Gamma(s)$ is the regularized gamma function and $\Gamma(s , z) = \int_{z}^{\infty} x^{s - 1} e^{ - x} \mathrm{d}x$ the incomplete gamma function.

We will be interested in some quantile $t_\alpha$ of the failure time distribution.
In the case of a continuous distribution function $F_T(t)$, the $\alpha$-quantile $t_\alpha$ satisfies $F_T(t_\alpha) = \alpha$, i.\,e., it represents the time up to which under normal use conditions $\alpha \cdot 100$ percent of the units fail and $(1 - \alpha) \cdot 100$ percent of the units persist. 
The distribution function and, hence, the quantile $t_\alpha = t_\alpha(\boldsymbol{\beta})$ depends on the parameter vector $\boldsymbol{\beta}$ in which the quantile $t_\alpha$ is a decreasing functions of the linear predictor $\beta_0 + \beta_1 x_u$. 

With this functional relationship the maximum likelihood estimator for the quantile $t_\alpha$ is given by $\widehat {t}_{\alpha} = t_{\alpha}(\widehat{\boldsymbol\beta})$, where $\widehat{\boldsymbol\beta}$ is the maximum likelihood estimator of ${\boldsymbol\beta}$.
The performance of these estimators is measured by their asymptotic variance $\mathrm{aVar}(\widehat {t}_{\alpha})$, and design optimization will be conducted with respect to the minimum asymptotic variance criterion, i.\,e.\ an optimal design minimizes $\mathrm{aVar}(\widehat {t}_{\alpha})$.
This criterion is commonly used in planning degradation tests when experimenters are interested in accurately estimating reliability properties of a system over its life cycle. 

If the distribution function $F_T(t)$ is strictly increasing with continuous density $f_T(t) = F_T^{\prime}(t)$, the asymptotic variance can be derived by the delta method from the information matrix in Section~\ref{sec:1.3} as
\begin{equation}
	\label{eq-avar-gamma-single}
		\mathrm{aVar}(\widehat {t}_{\alpha}) = \mathbf{c}^{T} \mathbf{M}_{\boldsymbol\beta}(\xi)^{-1} \mathbf{c} ,
\end{equation}
where $\mathbf{c} = \partial t_\alpha(\boldsymbol{\beta}) / \partial \boldsymbol{\beta}$ is the vector of partial derivatives of $t_\alpha = F_T^{-1}(\alpha)$ with respect to the components of the parameter vector $\boldsymbol{\beta}$ evaluated at the  true values of $\boldsymbol{\beta}$. 
Let $g(s) = Q(s, z_{0} / \nu)$ be the regularized gamma function with the second argument fixed to $z_{0} / \nu$, then $t_\alpha = g^{-1}(\alpha)/\gamma(x_u)$ by \eqref{eq-failure-time-distribution} and the vector $\mathbf{c}$ of partial derivatives can be written as $\mathbf{c} = - t_\alpha (1, x_u)^T$, where the minus sign and the scaling factor $t_\alpha$ do not affect the optimization problem.
Hence, the minimum asymptotic variance criterion is equivalent to a $c$-criterion with $\mathbf{c} = (1, x_u)^T$, i.\,e., extrapolation of the linear component $\beta_0 + \beta_1 x_u$ at the normal use condition $x_u$, and standard optimization methods for $c$-criteria can be employed. 
In particular, the design optimization does not depend on which quantile $t_\alpha$ is to be estimated, and the obtained design is simultaneously optimal for all $\alpha$.

Because the information matrix $\mathbf{M}_{\boldsymbol\beta}(\xi)$ depends on the parameter vector $\boldsymbol\beta$, this affects the design optimization. 
Hence, nominal values have to be assumed for these parameters, and locally optimal designs can be obtained for those nominal values.
Numerical calculations indicate that the locally optimal designs $\xi^*$ are supported on the endpoints of the design region $\mathcal{X}$, i.\,e., they are of the form $\xi^* = \xi_{w^*}$, where $\xi_w$ denotes a design with weight $w_1 = w$ on $x_1 = 0$ and weight $w_2 = 1 - w$ on $x_2 = 1$.
Under this premise the optimal weight $w^*$ can be determined analytically by Elfving's theorem \cite{elfving1952},
\begin{equation}
	\label{eq-w-opt-single}
w^* = \frac{(1 + |x_u|) \sqrt{\lambda(\beta_0 + \beta_1)}}{(1 + |x_u|) \sqrt{\lambda(\beta_0 + \beta_1)} + |x_u| \sqrt{\lambda(\beta_0)}}
\end{equation}
for (standardized) normal use condition $x_u < 0$.
This optimal weight $w^*$ is a decreasing function in the distance $|x_u|$ between the normal use condition and the lowest stress level $x_1 = 0$, and it decrease from $w^* = 1$ when formally letting $x_u = 0$ to $\sqrt{\lambda(\beta_0 + \beta_1)} / (\sqrt{\lambda(\beta_0 + \beta_1)} + \sqrt{\lambda(\beta_0)})$ for $x_u \to - \infty$, where this lower bound is larger than $0.5$ since $\beta_1 > 0$ and the intensity $\lambda(z)$ is an increasing function in $z$.

Concerning the parameters $\beta_0$ and $\beta_1$ the optimal weight $w^*$ is increasing in the slope parameter $\beta_1$ while it does not seem to be sensitive with respect to the intercept parameter $\beta_0$ as will be illustrated in Figure~\ref{w-vs-Bo} and Figure~\ref{w-vs-B1} below for some nominal values.
Therefore it is of interest to check how a misspecification of the nominal values for $\boldsymbol{\beta}$ may affect the performance of a locally optimal design $\xi^* = \xi_{w^*}$.
To measure the performance we make use of the concept of efficiency 
\begin{equation}
\label{eff-single-gamma}
	\mathrm{eff}_{\mathrm{aVar}}(\xi; \boldsymbol{\beta}) = \frac{\mathrm{aVar}_{\boldsymbol{\beta}}(\widehat{t}_{\alpha}; \xi_{\boldsymbol{\beta}}^*)}{\mathrm{aVar}_{\boldsymbol{\beta}}(\widehat{t}_{\alpha}; \xi)}
\end{equation}
of a design $\xi$ with respect to the asymptotic variance for estimating $t_\alpha$ when $\boldsymbol{\beta}$ is the true value of the parameter, where $\mathrm{aVar}_{\boldsymbol{\beta}}(\widehat{t}_{\alpha}; \xi)$ denotes the asymptotic variance of $\widehat{t}_{\alpha}$ at $\boldsymbol{\beta}$, when the design $\xi$ is used, and $\xi_{\boldsymbol{\beta}}^*$ is the locally optimal design at $\boldsymbol{\beta}$.
This efficiency attains a value between $0$ and $1$.
It can be interpreted as the proportion of units needed, when the locally optimal design $\xi_{\boldsymbol{\beta}}^*$ is used, to obtain the same precision in the asymptotic variance as for the design $\xi$ under consideration.
Thus high values of the efficiency are advantageous for a design to be used.

\subsection{Numerical example}
\label{sec:1.4}
In this example we consider an accelerated degradation experiment as described in Subsection~\ref{sec:1.2} with standardized normal use condition $x_u = - 0.4$, underlying gamma process with scale parameter $\nu = 1$ and degradation threshold $z_0 = 5.16$.
We will be interested in estimating the median $t_{0.5}$ of the failure time $T$ due to degradation.
The standardized observation times are $t_j = 0.25$, $0.5$, $0.75$ and $1$, i.\,e., there are $k = 4$ degradation increments measured on time intervals of constant length $\Delta = 0.25$.
With respect to  the location parameters we assume the nominal values $\beta_0 = 0.23$ for the intercept and $\beta_1 = 0.53$ for the slope.
For these parameter values, the distribution function $F_T(t)$ of the failure time $T$ is exhibited in Figure~\ref{fig-joint-failure-distrib-gammagamma} below as $F_{T_1}$.
The corresponding median failure time $t_{0.5} = 5.39$ for which $F_T(t_{0.5}) = 1/2$ is indicated in Figure~\ref{fig-joint-failure-distrib-gammagamma} by a dashed vertical line. 

To find the optimal design $\xi^*$ for estimating the median failure time $t_\alpha$, we apply the multiplicative algorithm following  \citep{torsney2009multiplicative} for the standardized stress parameter $x$ on a grid with increments of size $0.01$ on the design region $\mathcal{X} = [0,1]$. 
The optimal design $\xi^*$ is found to be of the form $\xi_w$ with optimal weight $w^*  = 0.79$ at the lowest stress level ($x = 0$) and weight $1 - w^* = 0.21$ at the highest stress level ($x = 1$) in the experiment.

\begin{figure}[!tbp]
  \centering
  \begin{minipage}[b]{0.4\textwidth}
   \centering
	\includegraphics[width=\textwidth]{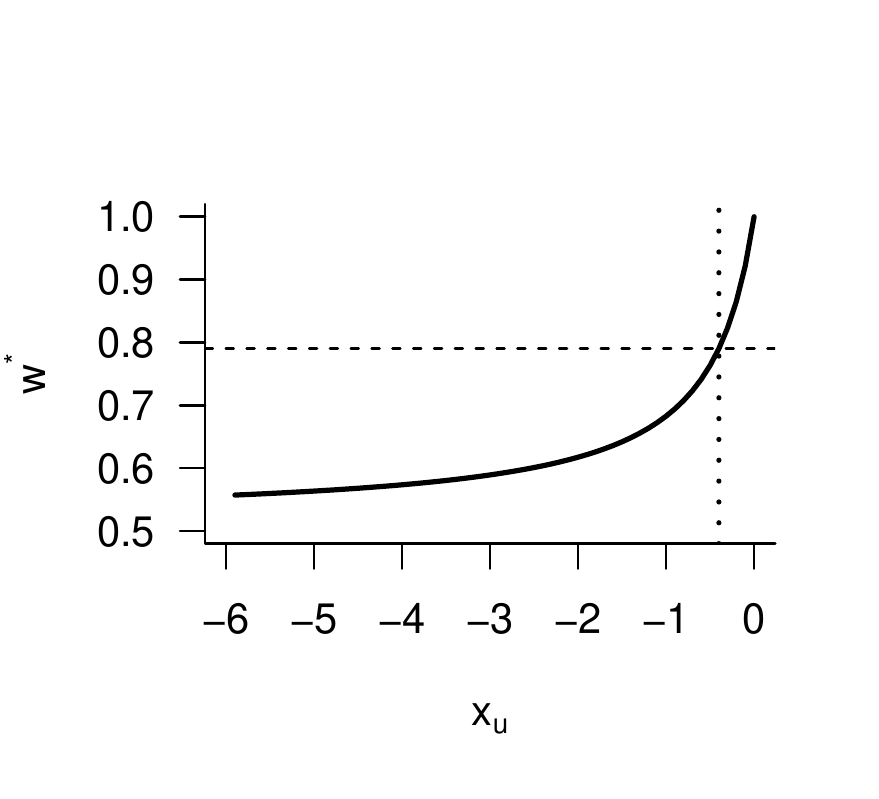}
	\caption{Optimal weights $w^*$ in dependence on the normal use condition $x_u$ for the univariate gamma process in the example of Subsection~\ref{sec:1.4}}
	\label{weight-univariate-gamma}
  \end{minipage}
  \hfill
 \begin{minipage}[b]{0.42\textwidth}
  \includegraphics[width=\textwidth]{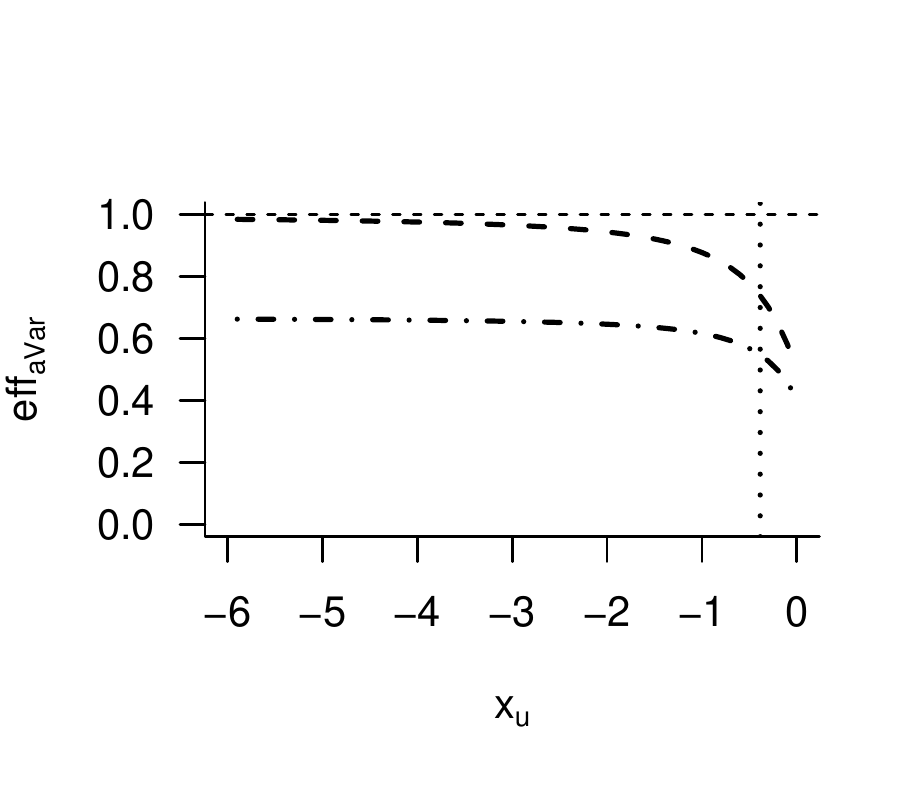}
	\caption{Efficiency of $\bar\xi_2$ (solid line) and $\bar\xi_3$ (dashed line) in dependence on the normal use condition $x_{u}$ in the example of Subsection~\ref{sec:1.4}}
	\label{eff-x1-gamma-singlee}
  \end{minipage}
\end{figure}

For illustrative purposes, the optimal weight $w^*$ is plotted in Figure~\ref{weight-univariate-gamma} as a function of the normal use condition $x_u$ when the nominal values of the other parameters are held fixed. 

For the normal use condition $x_u$ close to the lowest value $x = 0$ of the design region, the optimal weight $w^*$ approaches $1$, and $w^*$ decreases to $0.516$ when $x_u$ is far away from the design region ($x_u \to - \infty$).
 The nominal value $x_u = - 0.4$ and the corresponding optimal weight $w^* = 0.79$ are indicated in Figure~\ref{weight-univariate-gamma} by a vertical and a horizontal dashed line, respectively.
To imagine the gain in applying the optimal design $\xi^*$, we exhibit the efficiency  \eqref{eff-single-gamma} of commonly used standard designs $\bar\xi_2$ and $\bar\xi_3$ in Figure~\ref{eff-x1-gamma-singlee}, when the nominal values for the other parameters are held fixed.
In this comparison the designs $\bar\xi_m$ are uniform on $m$ equidistant stress values $x_1, ..., x_m$ covering the whole range of the design region $0 \leq x \leq 1$.
In particular, $\bar\xi_2$ is of the form $\xi_w$ with $w = 1/2$, and $\bar\xi_3$ assigns weight $1/3$ to each of the endpoints ($x = 0$) and ($x = 1$) and to the midpoint ($x = 0.5$) of the design region.

The nominal value $x_{u} = - 0.4$ of the normal use condition is indicated in Figure~\ref{eff-x1-gamma-singlee} by a vertical dotted line. 
The uniform two-point design $\bar\xi_2$ shows a high efficiency at values $x_u$ which are sufficiently far  from the standardized design region. 
This is in accordance with the similarity of the weights in $\bar\xi_2$ and in the optimal design $\xi^*$ for such values.
For $x_u$ close to the lowest experimental stress level $x=0$ the efficiency of $\bar\xi_2$ drops to $50\,\%$. 
The uniform three-point design $\bar\xi_3$ shows a much lower efficiency throughout.
At the nominal value $x_u = - 0.4$ the efficiency of the uniform two- and three-point designs $\bar\xi_2$ and $\bar\xi_3$ is $\mathrm{eff}_{\mathrm{aVar}}(\bar\xi_2; \boldsymbol{\beta}) = 75\,\%$ and $\mathrm{eff}_{\mathrm{aVar}}(\bar\xi_3; \boldsymbol{\beta}) = 55\,\%$, respectively.
That means that for the optimal design $\xi^*$ only 75 percent of units are required compared to the design $\bar\xi_2$ and $55$ percent of units compared to the design $\bar\xi_3$ to achieve the same accuracy for estimating the median failure time.

\begin{figure}[!tbp]
  \centering
  \begin{minipage}[b]{0.4\textwidth}
    \includegraphics[width=\textwidth]{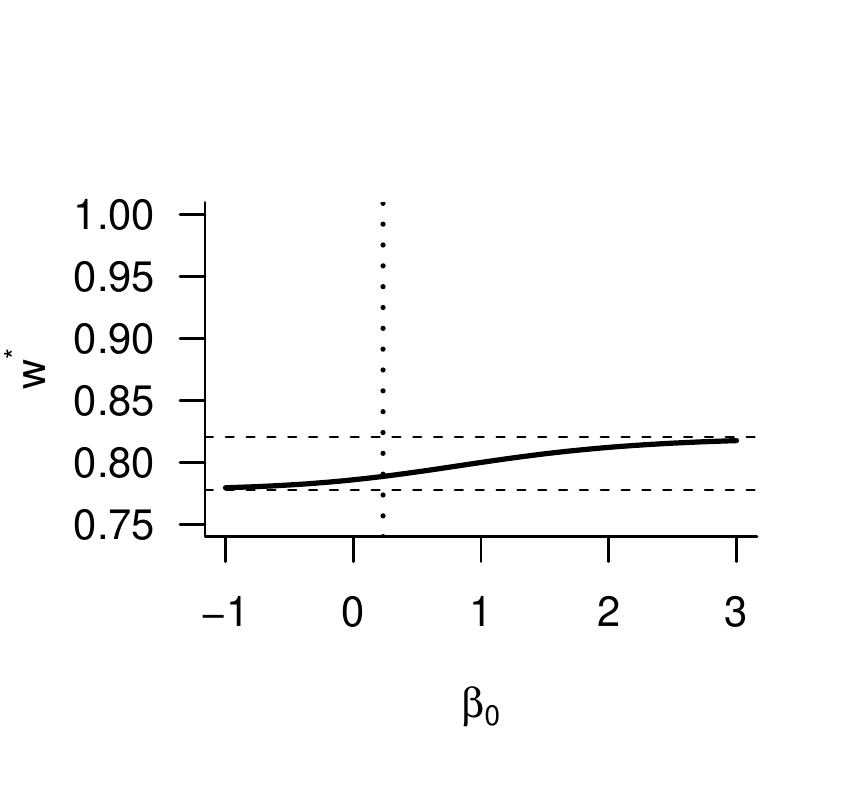}
    \caption{Dependence of the optimal weight $w^*$ on $\beta_{0}$ in the example of Subsection~\ref{sec:1.4}}
\label{w-vs-Bo}
  \end{minipage}
  \hfill
 \begin{minipage}[b]{0.4\textwidth}
    \includegraphics[width=\textwidth]{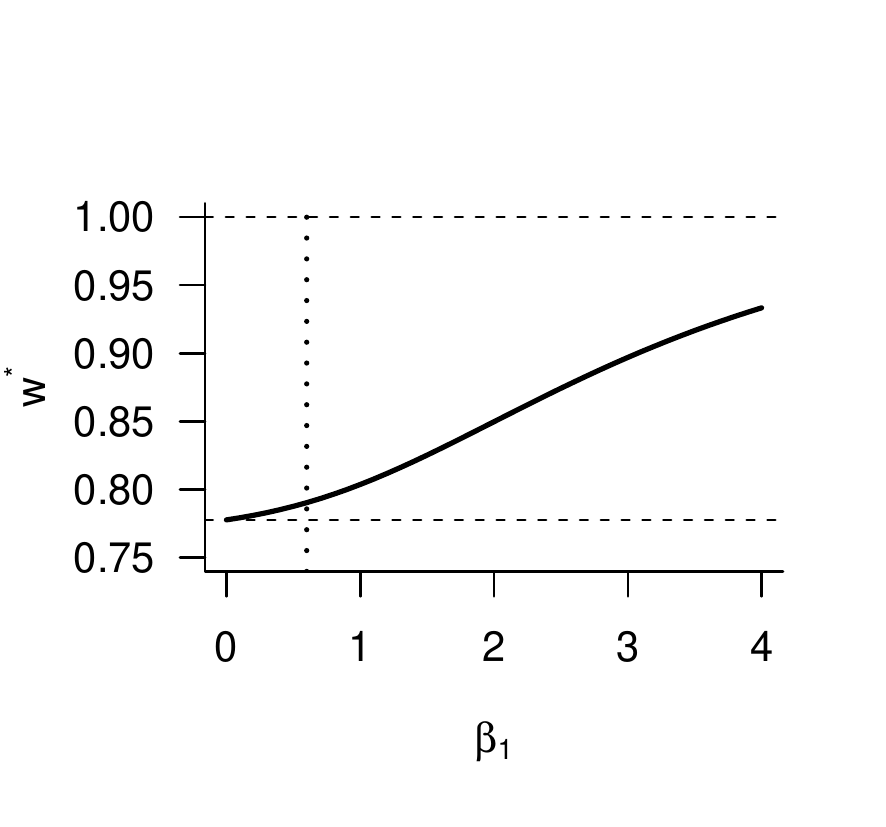}
    \caption{Dependence of the optimal weight $w^*$ on $\beta_{1}$ in the example of Subsection~\ref{sec:1.4}}
\label{w-vs-B1}
  \end{minipage}
\end{figure}

To assess the sensitivity of the locally optimal design $\xi^* = \xi_{w^*}$ we plot the optimal weights $w^*$ in dependence on the intercept and slope parameters $\beta_0$ and $\beta_1$ in Figure~\ref{w-vs-Bo} and  Figure~\ref{w-vs-B1}, respectively, while the other nominal values are held fixed. As mentioned earlier, the 
\begin{figure}[!tbp]
  \centering
 \begin{minipage}[b]{0.4\textwidth}
    \includegraphics[width=\textwidth]{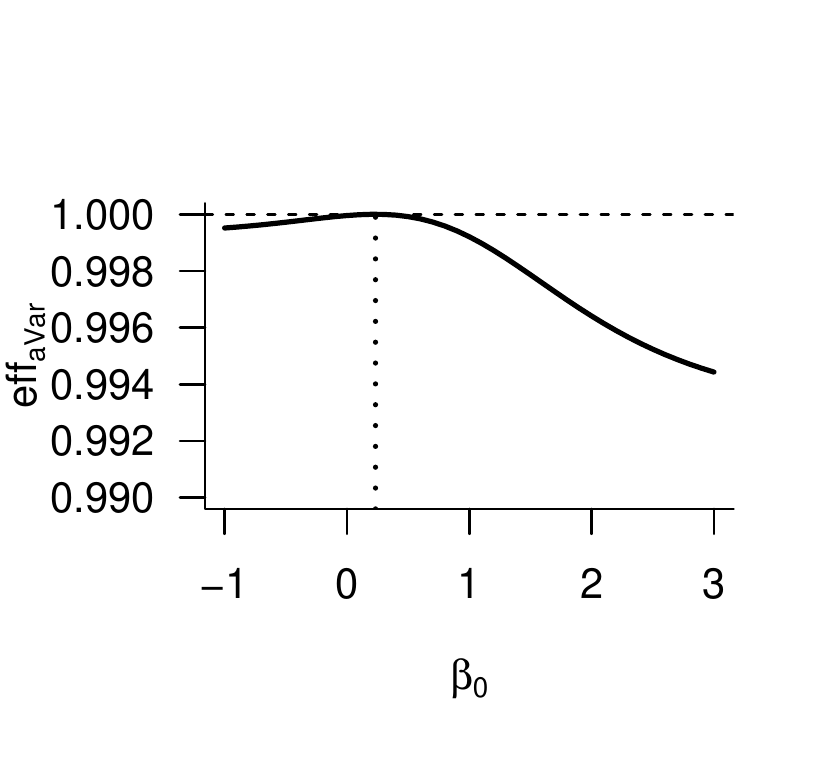}
   \caption{Efficiency of $\xi^*$ in dependence on $\beta_{0}$ in the example of Subsection~\ref{sec:1.4}}
\label{effvsB0gammasingle}
  \end{minipage}
  \hfill
  \begin{minipage}[b]{0.4\textwidth}
    \includegraphics[width=\textwidth]{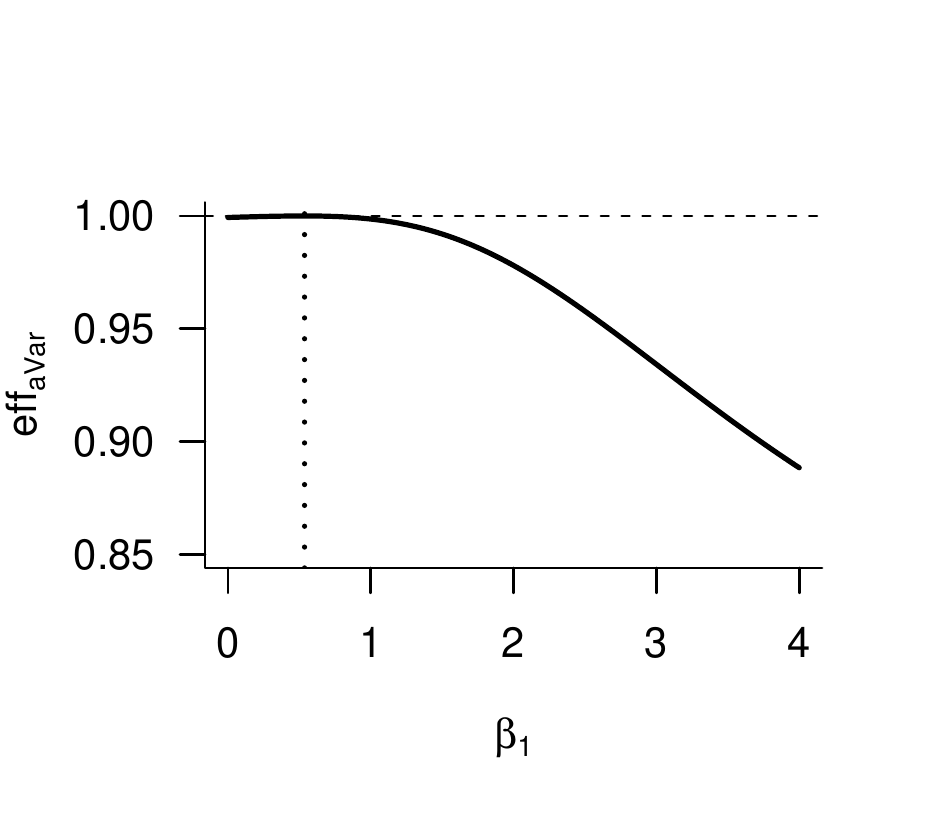}
   \caption{Efficiency of $\xi^*$ in dependence on $\beta_{1}$ in the example of Subsection~\ref{sec:1.4}}
\label{effvsB1gammasingle}
  \end{minipage}
\end{figure}

Finally, to judge the performance of the locally optimal design $\xi^* = \xi_{\boldsymbol{\beta}}^*$ at the proposed nominal values $\beta_0 = 0.23$ and $\beta_1 = 0.53$ of the location parameters under misspecifications, we show the efficiency in dependence on the intercept and slope parameters $\beta_0$ and $\beta_1$ in Figure~\ref{effvsB0gammasingle} and  Figure~\ref{effvsB1gammasingle}, respectively, while the other parameters are held fixed to their nominal values. Figure~\ref{effvsB0gammasingle} displays that the optimal design $\xi^*$ maintain its efficiency under misspecifications of $\beta_0$. In contrast, Figure~\ref{effvsB1gammasingle} depicts that misspecifications of the slope $\beta_1$ substantially affects the efficiency of the design $\xi^*$ and more attention should be paid to a correct specification of the nominal value for the slope parameter opposite to the intercept.

\section{Bivariate accelerated degradation testing with two gamma processes}
\label{sec:222}
We consider now the optimal design problem for a bivariate degradation process incorporating serially two independent failure modes which means that a failure of the system occurs when one of the two components fail.

\subsection{Model formulation}
\label{formulation-bivariate-Gamma}
We assume that in the two degradation components takes place according to independent gamma processes $Z_{1t}$ and $Z_{2t}$, respectively, as described in Section~\ref{sec:11}, where for both processes $Z_{lt}$ the rate $\gamma_l(x) = e^{\beta_{l0} + \beta_{l1} x}$ depends on the same standardized accelerating stress variable $x \in \mathcal{X} = [0,1]$ via a linear trend $\beta_{l0} + \beta_{l1} x$ under the log link as in \eqref{shape}. 
By assumption the degradation increments $Y_{ilj} = Z_{lt_j} - Z_{lt_{j-1}}$ of both components during the $j$th time interval of length $\Delta_j$ are all gamma distributed with shape $\gamma_l(x_i) \Delta_{j}$ and scale $\nu_l$, $l = 1, 2$, and independent.

The failure times $T_1$ and $T_2$ of the components for soft failure due to degradation are defined as in Subsection~\ref{sec:1.345}.
The failure of the system occurs when either of the two components fail, and the failure time $T$ of the system is defined by $T = \min\{T_1, T_2\}$.
Because of the independence of the underlying processes, the failure times $T_1$ and $T_2$ of the components are independent.

\subsection{Information}
\label{sec:2.3}
Denote by $\boldsymbol{\beta}_l=({\beta}_{l1},\beta_{l2})^T$ the marginal parameter vector associated with the $l$th failure mode.
Because of the independence of the components the joint log-likelihood of $\boldsymbol{\beta}_1$ and $\boldsymbol{\beta}_l$ is the sum $\ell(\boldsymbol{\beta}_1, \ell(\boldsymbol{\beta}_2; y_{111}, ..., y_{n2k}) = \ell(\boldsymbol{\beta}_1; y_{111}, ..., y_{n1k}) + \ell(\boldsymbol{\beta}_2; y_{121}, ..., y_{n2k})$ of the log-likelihoods $\ell(\boldsymbol{\beta}_l; y_{1l1}, ..., y_{nlk})$ of the components given by \ref{eq-loglik-single}. 
Hence, the maximum likelihood estimators $\widehat{\boldsymbol{\beta}}_l$ of $\boldsymbol{\beta}_l$ in the whole system coincides with those in the marginal models and the joint information matrix $\mathbf{M}_{\boldsymbol{\beta}_1, \boldsymbol{\beta}_2}(x_1, ..., x_n)$ for all parameters is block diagonal, $\mathbf{M}_{\boldsymbol{\beta}_1, \boldsymbol{\beta}_2}(x_1, ..., x_n) = \left( \begin{array}{cc} \mathbf{M}_{\boldsymbol{\beta}_1}(x_1, ..., x_n) & \mathbf{0} \\ \mathbf{0} & \mathbf{M}_{\boldsymbol{\beta}_2}(x_1, ..., x_n) \end{array}\right)$, where the diagonal blocks $\mathbf{M}_{\boldsymbol{\beta}_l}$ are the marginal information matrices for the single failure modes as specified in Subsection~\ref{sec:1.3}. 
Accordingly, for approximate designs $\xi$ the standardized information matrix 
\begin{equation}
\label{info-stand-block}
\mathbf{M}_{\boldsymbol{\beta}_1, \boldsymbol{\beta}_2}(\xi) = \left(
		\begin{array}{cc} 
			\mathbf{M}_{\boldsymbol{\beta}_1}(\xi) & \mathbf{0} 
			\\ 
			\mathbf{0} & \mathbf{M}_{\boldsymbol{\beta}_2}(\xi) 
		\end{array} 
	\right).
\end{equation}
is also block diagonal with the marginal information matrices $\mathbf{M}_{\boldsymbol{\beta}_l}(\xi)$ on the diagonal.

\subsection{Optimality criterion based on the failure time distribution}
\label{sec:1.222345}
As in section~\ref{sec:1.345} we are interested in characteristics of the failure time distribution of soft failure due to degradation under normal use condition ${x}_u$. 
The marginal failure times $T_l$ under normal use condition are defined as the first time $t$ the degradation path $Z_{u,lt}$ reaches or exceeds the corresponding threshold $z_{l0}$, i.\,e., $T_l = \inf \{t \geq 0;\, Z_{u,lt} \geq z_{l0}\}$.
A failure of the system occurs if one of the components fail.
Hence, the failure time $T$ of the system is defined by $T = \mathrm{min}\{T_1, T_2\}$. 
Because of the independence of the components the survival function $1 - F_T(t) = \mathrm{P} (T_1 > t, T_2 > t)$ factorizes into the marginal survival functions $1 - F_{T_l}(t)$.
Hence, the failure time distribution of the system can be expressed as
\begin{equation}
\label{joint-failure-distrib-gamma-gamma}
		F_T(t) = 1 - (1 - F_{T_1}(t)) (1 - F_{T_2}(t)) ,
\end{equation}
where $F_{T_l}(t) = Q(\gamma_l(x_{u}) t, z_{l0} / \nu_l)$ by \eqref{eq-failure-time-distribution}.

As in Subsection~\ref{sec:1.345}, we will consider quantiles $t_\alpha$ of the failure time distribution. 
Also here the distribution function $F_T$ and, hence, the quantile $t_\alpha = t_\alpha(\boldsymbol{\beta}_1, \boldsymbol{\beta}_2)$ is a function of the parameters and the maximum likelihood estimate $\widehat{t}_\alpha = t_\alpha(\widehat{\boldsymbol{\beta}}_1, \widehat{\boldsymbol{\beta}}_2)$ of the quantile $t_\alpha$ is based on the maximum likelihood estimates $\widehat{\boldsymbol{\beta}}_l$ of $\boldsymbol{\beta}$ for the components. 

The task of designing the experiment is to provide an as precise estimate of the $\alpha$-quantile as possible, i.\,e., to minimize the asymptotic variance $\mathrm{aVar}(\widehat{t}_\alpha)$ of $\widehat{t}_\alpha$ at the normal use condition. 
As in Subsection~\ref{sec:1.345} the asymptotic variance can be obtained as $\mathrm{aVar}(\widehat {t}_{\alpha}) = \mathbf{c}^{T} \mathbf{M}_{\boldsymbol{\beta}_1, \boldsymbol{\beta}_2}(\xi)^{-1} \mathbf{c}$, where $\mathbf{c} = (\mathbf{c}_1^T, \mathbf{c}_2^T)^T$ and $\mathbf{c}_l = \partial t_\alpha(\boldsymbol{\beta}_1, \boldsymbol{\beta}_2) / \partial \boldsymbol{\beta}_l$ is the vector of partial derivatives of $t_\alpha$ with respect to the parameter vector $\boldsymbol{\beta}_l$ evaluated at the  true values of $\boldsymbol{\beta}_l$. 
Differently from the univariate case in Subsection~\ref{sec:1.345} there is no explicit formula for $t_{\alpha}$.
Therefore, the gradient vectors $\mathbf{c}_l$ will be derived  by the implicit function theorem as
\begin{equation}
	\label{iftheorem222}
	\frac{\partial t_{\alpha}}{\partial{\boldsymbol{\beta}}_l} = - \frac{1}{f_T(t_\alpha)} \, \frac{\partial  F_T(t_\alpha)}{\partial {\boldsymbol{\beta}}_l} 
\end{equation}
in terms of the failure time distribution $F_T(t)$, where $f_T(t) =  \partial  F_T(t) / \partial t$ is the density of $T$.
The common scaling factor $ c_0 = - 1 / f_T(t_\alpha)$ is irrelevant for the optimization problem.
Hence, the components of the $c$-criterion vector $\mathbf{c} = (\mathbf{c}_1^T, \mathbf{c}_2^T)^T$ can be reduced to $\mathbf{c}_l = \partial  F_T(t_{\alpha}) / \partial \boldsymbol\beta_l$. 
Based on equation \ref{joint-failure-distrib-gamma-gamma}, the gradient vectors $\mathbf{c}_l$ can be expressed as $\mathbf{c}_l = c_l (1, x_u)^T$ similar to the univariate case, where the constant $c_l$ can be expressed as, see \citep{tsai2012optimal},
\begin{equation}
	\begin{split}
		\label{gradient_cb1}
		c_l = & \kappa_l \left(1 - F_{T_{l^\prime}}(t_\alpha)\right) \left(\frac{\Gamma(\kappa_l)}{\Gamma(\kappa_l + 1)^2} (z_{l0} / \nu_l)^{\kappa_l}\, {}_{2}F_2(\kappa_l, \kappa_l; \kappa_l + 1, \kappa_l + 1; - z_{l0} / \nu_l) \right.
		\\
		& \left. \mbox{} + \left(Q(\kappa_l, z_{l0} / \nu_l) - 1\right) \left(\ln(z_{l0} / \nu_l) - \psi(\kappa_l)\right)\right)
	\end{split}
\end{equation}
is a positive constant depending on $\boldsymbol\beta_1$ and $\boldsymbol\beta_2$, 
 $\kappa_l = \gamma_l(x_{u}) t_\alpha$ is the shape parameter for an increment of the $l$th marginal process during time $t_\alpha$, ${}_{2}F_2$ denotes the generalized hypergeometric function
\[
	{}_{2}F_2(\alpha, \alpha; \alpha + 1, \alpha + 1; - z) = 
	1 + \sum_{k = 1}^\infty \left(\frac{\alpha}{\alpha + k}\right)^2 \frac{(- z)^k}{k!} ,
\]
and $l^\prime$ is the index of the respective other component, i.\,e., $l^\prime = 2$ if $l = 1$ and vice versa.

Since the information matrix in \eqref{info-stand-block} is block-diagonal, the optimality criterion   
\begin{equation} 
	\label{111fghghfghgammagamma}
	\mathrm{aVar}(\widehat {t}_{\alpha}) 
	 =
	c_0^2 (c_1^2 (1, x_u) \mathbf{{M}}_{{\boldsymbol\beta}_1}(\xi)^{-1} (1, x_u)^T + c_2^2 (1, x_u) \mathbf{{M}}_{{\boldsymbol\beta}_2}(\xi)^{-1} (1, x_u)^T)
\end{equation}
is a weighted sum of the optimality criteria for the single components stated in Subsection~\ref{sec:1.345} and constitutes, hence, a compound criterion.

In the special case that the nominal values are identical for both components, i.\,e., $\boldsymbol{\beta}_1 = \boldsymbol{\beta}_2$, the optimal design $\xi^*$ for a single component will also be optimal for the bivariate failure process, independent of $\alpha$.
In general, however, the optimal design for the bivariate failure process has to be a compromise of the marginal optimal designs for the components.

\subsection{Numerical example}
\label{sec:2.5gammagamma}
In this example we consider an accelerated degradation experiment with two failure components following two independent gamma processes.
The first process is specified as in Subsection~\ref{sec:1.4} with scale parameter $\nu_1 = 1$, degradation threshold $z_{10} = 5.16$ and nominal values $\beta_{10} = 0.23$ for the intercept and $\beta_{11} = 0.53$ for the slope.
For the second process we assume a scale parameter $\nu_2 = 0.88$, a degradation threshold $z_{20} = 4.60$ and nominal values $\beta_{20} = 0.31$ for the intercept and $\beta_{21} = 0.35$ for the slope.
As in Subsection~\ref{sec:1.4} the standardized normal use condition is $x_u = - 0.40$ and the processes are measured at $k = 4$ standardized time points $t_j = 0.25$, $0.5$, $0.75$ and $1$ with time intervals of constant length $\Delta = 0.25$.
Also here we will be interested in estimating the median failure time $t_{0.5}$.

\begin{figure}
  \centering
 \begin{minipage}[b]{0.4\textwidth}
    \includegraphics[width=\textwidth]{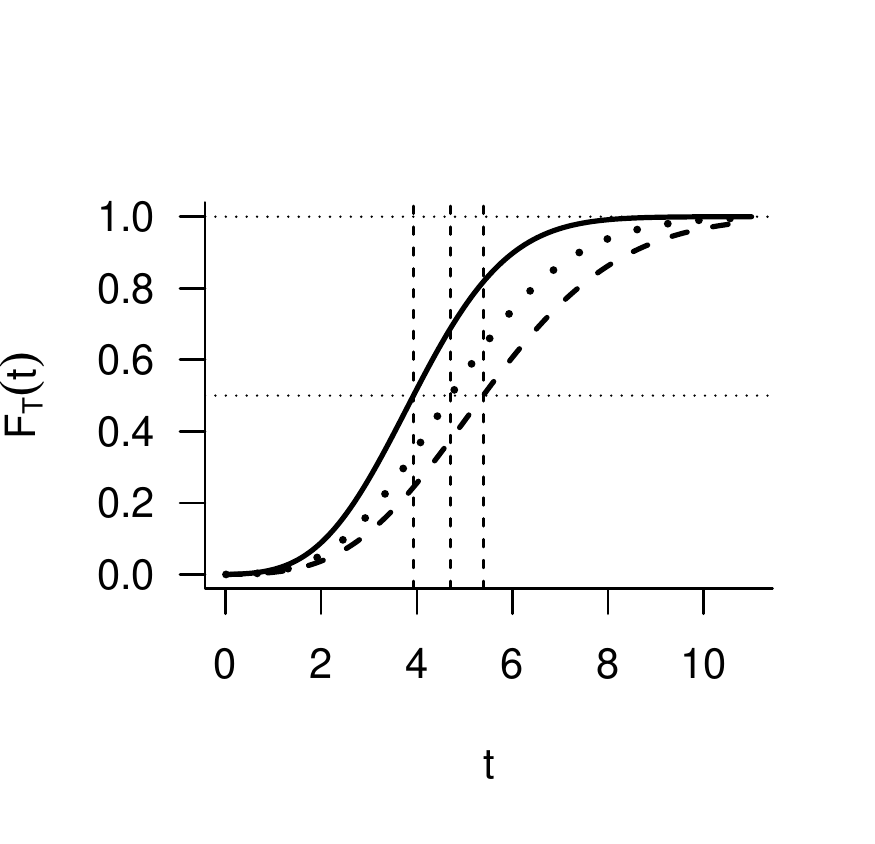}
	\caption{Failure time distributions $F_T(t)$ (solid line), $F_{T_1}(t)$ (dashed line), and $F_{T_2}(t)$ (dotted line) for the bivariate gamma process in the example of Subsection~\ref{sec:2.5gammagamma}} 
	\label{fig-joint-failure-distrib-gammagamma}
  \end{minipage}
  \hfill
  \begin{minipage}[b]{0.4\textwidth}
    \includegraphics[width=\textwidth]{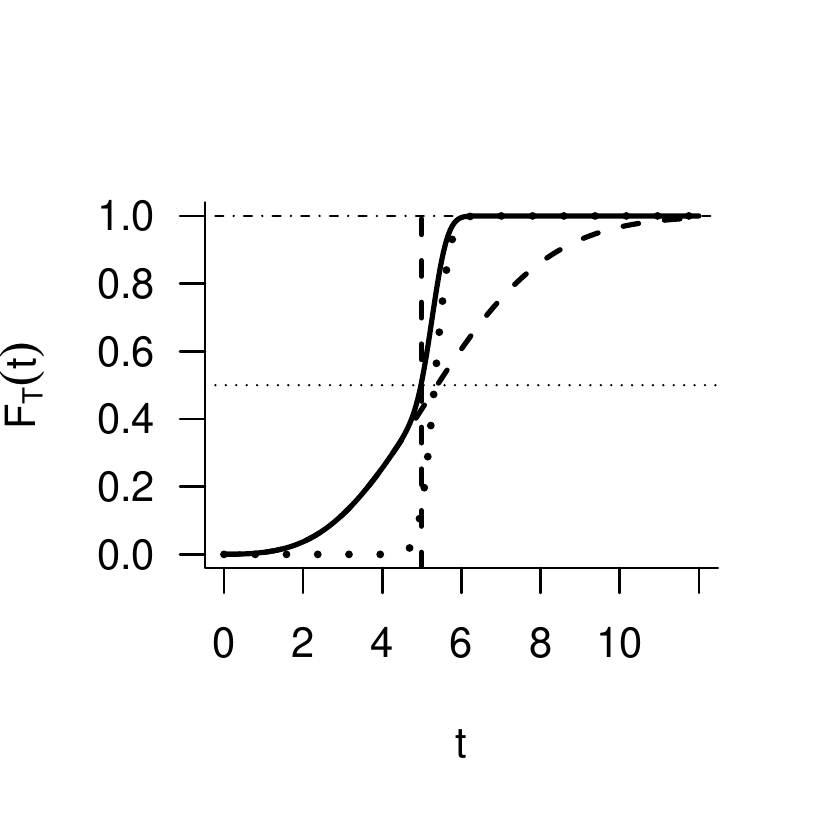}
	\caption{Failure time distributions $F_T(t)$ (solid line), $F_{T_1}(t)$ (dashed line), and $F_{T_2}(t)$ (dotted line) for the bivariate model with a gamma process ($T_1$)  and a linear mixed effect ($T_2$) component for the example of Subsection~\ref{sec:2.5}}
	\label{fig-failuretime-gammalinear}
  \end{minipage}
\end{figure}

The distribution function $F_T(t)$ of the combined failure time $T$ given by  \eqref{joint-failure-distrib-gamma-gamma} is plotted in Figure~\ref{fig-joint-failure-distrib-gammagamma} together with the distribution functions $F_{T_1}(t)$ and $F_{T_2}(t)$ of the failure times $T_1$ and $T_2$ in the components. 
The median failure time $t_{0.5} = 3.93$ satisfying $F_T(t_{0.5}) = 1/2$ is indicated there together with the median failure times for the single components by dashed vertical lines.

For estimating the median failure time, also here the optimal design is sought numerically by means of the multiplicative algorithm on an equidistant grid of step size $0.01$ on the design region. 
The locally optimal design obtained is of the form $\xi^* = \xi_{w^*}$ assigning optimal weights $w^* = 0.78$ to the lowest stress level $x = 0$ and $1 - w^* = 0.22$ to the highest stress level $x = 1$ in the design region.
This optimal weight is close to the solution for the first component (see Subsection~\ref{sec:1.4}) and shows a similar behavior, when the value ot the normal use condition is altered.
Similar considerations hold for the sensitivity with respect to misspecifications of the nominal values of the parameters.

\section{Bivariate accelerated degradation testing with a gamma process and a  linear mixed model}
\label{sec:2} 
In section \ref{sec:222} we considered a degradation process with two response components where each is modeled by a gamma model. 
In this section we consider a bivariate process with two different degradation models.
The first degradation mode is modeled by a gamma process as in Section~\ref{sec:11}.
As in Section~\ref{sec:222} the degradation increments of this component are denoted by $Y_{i1j}$ for unit $i$ during a time intervals of length $\delta_j = t_j - t_{j - 1}$, $j = 1, ..., k$.
The second degradation mode is given by a linear model with random intercept which will be described in the subsequent subsection and is a special case of the model treated in \citep{shat2021experimental}. 
Both failure modes are influenced by the same standardized accelerating stress variable $x \in \mathcal{X} = [0,1]$. 
Apart from that the degradation modes are assumed to be independent and, hence, do not have an interactive effect. 
As before, also here a failure of the system occurs when at least one of the marginal degradation paths exceeds its corresponding failure threshold.

\subsection{Model formulation of the second degradation component: Linear mixed model}
\label{sec:2.1}
Here we consider a linear regression model, similar to the model presented by \citep{doi:10.1002/asmb.2061}, for a single stress variable $x$. 
Measurements $Y_{i2j}$ of the second component at unit $i$ are taken at the same time points $t_1, ..., t_k$ as for the first component and additionally at the beginning of the degradation experiment, $t_0 = 0$, $j = 0, ..., k$.
These measurements are described by a hierarchical model. 
For each unit $i$ the observation $Y_{i2j}$ at time point $t_j$ is given by
\begin{equation} 
\label{modelindividualresponse}
Y_{i2j} = \mu_{i}(x_i, t_j) + \varepsilon_{i j} ,
\end{equation}
where $\mu_{i}(x, t)$ is the mean degradation path of the second marginal response of unit $i$ at time $t$, when stress $x$ is applied to unit $i$, and $\varepsilon_{i j}$ is the associated measurement error at time point $t_j$.
The mean degradation $\mu_{i}(x, t)$ is given by a linear model equation in the stress variable $x$ and in time $t$ with stress-time interaction,
\begin{equation} 
	\label{mean-degradation-unit}
	\mu_{i}(x, t) = \beta_{i20} + \beta_{21} x + \beta_{22} t + \beta_{23} x t  
\end{equation}
where only the intercept is unit specific and the  time and stress effects are the same for all units.
Hence, the response is given by
\begin{equation} 
	\label{LMEMtrztr765765756111}
	Y_{i2j} =  \beta_{i20} + \beta_{21} x_i + \beta_{22} t_j + \beta_{23} x_i t_j   + \varepsilon_{i j}.
\end{equation}
The measurement error $\varepsilon_{i j}$ is assumed to be normally distributed with zero mean and a time independent error variance $\sigma_\varepsilon^2 > 0$.
Moreover, the error terms are assumed to be independent within a unit over time. 

On the aggregate level it is assumed that the units are representatives of a larger entity.
The unit specific intercept $\beta_{i20}$ is modeled as a random effect, i.\,e., $\beta_{i20}$ is normally distributed with mean $\beta_{20}$ and variance $\sigma_0^2 > 0$.
All measurement errors $\varepsilon_{ij}$ and random effects $\beta_{i20}$ are assumed to be independent.
For transferring the results, it is assumed that the model defined in equation (\ref{LMEMtrztr765765756111}) also holds for units under normal use condition $x_{u}$.

\subsection{Information for the second degradation component: Linear mixed model}
\label{sec-linear-information}
To derive the information matrix in the mixed effects model we first write the model in vector notation. 
Denote by $\boldsymbol{\beta}_2 = (\beta_{20}, \beta_{2 1}, \beta_{2 2}, \beta_{2 3})^T$ the vector of fixed effect (aggregate) location parameters and by $\boldsymbol{\varsigma} = (\sigma_0^2, \sigma_{\varepsilon}^2)^T$ the vector of variance parameters.
The $(k + 1)$-dimensional vector of observations $\mathbf{Y}_{i2} = (Y_{i20}, ..., Y_{i2k})^T$ at unit $i$ is multivariate normal with expectation $\mathrm{E}(\mathbf{Y}_{i2}) = \left(\mathbf{D} \otimes (1, x_i)\right) \boldsymbol{\beta}_2$, where $\mathbf{D} = \left((1, t_0)^T, ..., (1, t_k)^T\right)^T$ is the ``design'' matrix for the time variable and ``$\otimes$'' denotes the Kronecker product, and compound symmetric covariance matrix $\mathrm{Cov}(\mathbf{Y}_{i2}) = \mathbf{V}$ with diagonal  entries $\sigma_0^2 + \sigma_\varepsilon^2$ and off-diagonals $\sigma_0^2$.
The elemental information matrix (per unit) $\mathbf{M}_{{\boldsymbol\beta}_2, \boldsymbol{\varsigma}}(x_{i}) = \left( \begin{array}{cc} \mathbf{M}_{{\boldsymbol\beta}_2}(x_{i}) & \mathbf{0} \\ \mathbf{0} & \mathbf{M}_{\boldsymbol{\varsigma}} \end{array} \right)$ of the linear mixed model component is block diagonal with the elemental information matrices $\mathbf{M}_{{\boldsymbol\beta}_2}(x) = \mathbf{D}^T \mathbf{V}^{-1} \mathbf{D} \otimes \left( \begin{array}{cc} 1 & x \\ x & x^2 \end{array} \right)$ for the location parameters and $\mathbf{M}_{\boldsymbol{\varsigma}}$ for the variance parameters on the diagonal, where $\mathbf{M}_{\boldsymbol{\varsigma}}$ does not depend on the setting $x$ of the stress variable.

Accordingly, also for an approximate design $\xi$, the standardized information matrix 
\begin{equation}
	\label{eq-info-linear-mixed-approx}
	\mathbf{M}_{{\boldsymbol\beta}_2, \boldsymbol{\varsigma}}(\xi) 
	= \left( 
		\begin{array}{cc}
			\mathbf{D}^T \mathbf{V}^{-1} \mathbf{D} \otimes \mathbf{M}(\xi) & \mathbf{0}
			\\
			\mathbf{0} &\mathbf{M}_{\boldsymbol\varsigma}
		\end{array}
	\right) 
\end{equation}
 of the second component is block diagonal, where $\mathbf{M}(\xi) = \sum_{i = 1}^{m} w_i \left( \begin{array}{cc} 1 & x_i \\ x_i & x_i^2 \end{array}  \right)$is the standardized information matrix of linear fixed effect regression model which does not depend on the parameters.
For further details of the linear mixed model see \citep{shat2021experimental}.

\subsection{Failure time distribution for the second degradation component: Linear mixed model}
\label{sec-failure-time-mixed-model}
As mentioned in section \ref{sec:1.345} we are interested in characteristics of the failure time distribution of soft failure due to degradation.
Therefore it is assumed that the model equation $\mu_{u}(t) = \beta_{u20} + \beta_{21} x_u + \beta_{22} t + \beta_{23} x_u t$ is also valid under the normal use condition, where $\mu_{u}$ denotes the mean degradation path for a unit ``$u$'' under the normal use condition $x_u$ and $\beta_{u20}$ is the random intercept of $u$.
We further denote by $\mu(t) = \mathrm{E}(\mu_{u}(t)) = \beta_{20} + \beta_{21} x_u + \beta_{22} t + \beta_{23} x_u t$ the aggregate degradation path under normal use condition.

A soft failure due to degradation for the second response component is defined as the exceedance of the degradation over a  failure threshold $y_{20}$.
This definition is based on the mean degradation path $\mu_u(t)$ and not on a ``real'' path subject to measurement errors.
The failure time $T_2$ under normal use condition is then defined as the first time $t$ the mean degradation path $\mu_{u}(t)$ reaches or exceeds the threshold $y_{20}$, i.\,e.\ $T_2 = \min \{t \geq 0;\, \mu_{u}(t) \geq y_{20}\}$.
Because the random intercept $\beta_{u20}$ is involved in the mean degradation path, the failure time $T_2$ is random.

As in the previous sections, we will describe the characteristics of the failure time $T_2$ by its distribution function $F_{T_2}(t)$.
We note that $T_2 \leq t$ if and only if $\mu_{u}(t) \geq y_{20}$ and, hence, we can derive
\begin{equation}
\label{lmem22dfgdfgfdg22}
   F_{T_2}(t) = \mathrm{P}(\mu_{u}(t) \geq y_{20}) = \Phi((\mu(t) - y_{20}) / \sigma_0) ,
\end{equation}
where $\Phi$ denotes the distribution function of the standard normal distribution.
For later use we also state the gradient 
\begin{equation}
	\label{eq-grad-linear}
	\partial F_{T_2}(t) / \partial \boldsymbol{\beta}_2 = \sigma_0^{-1} \varphi\left((\mu(t_\alpha) - y_{20}) / \sigma_0 \right) (1, t)^T \otimes (1, x_u)^T 
\end{equation}
of $F_{T_2}(t)$ with respect to the vector $\boldsymbol{\beta}_2$ of location parameters (cf.\ \citep{shat2021experimental}), where $\varphi$ denotes the density of the standard normal distribution.

\subsection{Estimation and information in the combined model}
\label{sec:3.3}
The combined model parameters $\boldsymbol{\beta}_1$, $\boldsymbol{\beta}_2$ and $\boldsymbol{\varsigma}$ can be estimated by means of the maximum likelihood method.
As stated in Subsection~\ref{sec:1.222345}, in the combined model the maximum likelihood estimates coincide with those for the single components because of the independence between the failure modes.
Accordingly, the combined information matrix for all parameters is block diagonal with the information matrices for the components on the diagonal.
In view of \eqref{eq-info-linear-mixed-approx} the information matrix of an approximate design $\xi$ is given by
\begin{equation}
	\label{eq-info-combined-approx}
	\mathbf{M}_{\boldsymbol{\beta}_1, \boldsymbol{\beta}_2, \boldsymbol{\varsigma}}(\xi) 
	= \left( 
		\begin{array}{ccc}
			\mathbf{M}_{{\boldsymbol\beta}_1}(\xi) & \mathbf{0} & \mathbf{0}
			\\
			\mathbf{0} & \mathbf{M}_{{\boldsymbol\beta}_2}(\xi) & \mathbf{0}
			\\
			\mathbf{0} & \mathbf{0} & \mathbf{M}_{\boldsymbol\varsigma}(\xi) 
		\end{array}
	\right) ,
\end{equation} 
where $\mathbf{M}_{{\boldsymbol\beta}_2}(\xi) = \mathbf{D}^T \mathbf{V}^{-1} \mathbf{D} \otimes \mathbf{M}(\xi)$ and $\mathbf{{M}}_{{\boldsymbol\beta}_1}(\xi)$ as in Subsection~\ref{sec:1.3}.

\subsection{Optimality criterion based on the joint failure time}
\label{sec:2.4}
The combined failure time $T$ is defined as the minimum of the marginal  failure times $T_1$ and $T_2$ for the single components derived in Subsections~\ref{sec:1.345} and \ref{sec-failure-time-mixed-model}. 
As in Subsection~\ref{sec:2.4}, the survival function of the joint failure time $T$ factorizes and, hence, the distribution function $F_T(t)$ can be expressed as $F_T(t)  = 1 - (1 - F_{T_1}(t)) (1 - F_{T_2}(t))$.
The quantiles $t_\alpha = t_\alpha(\boldsymbol\beta_1, \boldsymbol\beta_2, \boldsymbol\varsigma)$ are functions of both the location parameters $\boldsymbol\beta_1$ and $\boldsymbol\beta_2$ as well as on the variance parameters $\boldsymbol\varsigma$, in general.
Consequently, the maximum likelihood estimate of a quantile $t_\alpha$ is given by $\widehat{t}_\alpha = t_\alpha(\widehat{\boldsymbol\beta}_1, \widehat{\boldsymbol\beta_2}, \widehat{\boldsymbol\varsigma})$ in terms of the maximum likelihood estimates $\widehat{\boldsymbol\beta}_1$, $\widehat{\boldsymbol\beta}_2$ and $\widehat{\boldsymbol\varsigma}$ of the parameters ${\boldsymbol\beta}_1$, ${\boldsymbol\beta}_2$ and ${\boldsymbol\varsigma}$ in the components.
The asymptotic variance of $\widehat{t}_\alpha$ can again be obtained by the delta method and the implicit function theorem. 
By the block diagonal structure of the information matrix and the decomposition of the distribution function of the failure time we get
\begin{equation} 
	\label{eq-avar-gamma-linear}
	\mathrm{aVar}(\widehat {t}_{\alpha}) 
	=
	f_T(t_\alpha)^{-2} \left(c_1^2 \, (1, x_u) \mathbf{{M}}_{{\boldsymbol\beta}_1}(\xi)^{-1} (1, x_u)^T + c_2^2 \, (1, x_u) \mathbf{M}(\xi)^{-1} (1, x_u)^T + c_{\boldsymbol{\varsigma}}^2\right) ,
\end{equation}
where $f_T(t)$ is the density of $T$, $c_1$ is defined as in \eqref{gradient_cb1} with the distribution function $F_{T_2}(t_\alpha) = \Phi((\mu(t_\alpha) - y_{20}) / \sigma_0)$ of the linear mixed effect model inserted,
\begin{equation}
	\label{eq-c2-linear}
	c_2 = \left(1 - Q\left(\gamma(x_u) t_\alpha, z_{10} / \nu_1\right)\right) \sigma_0^{-1} \varphi\left((\mu(t_\alpha) - y_{20}) / \sigma_0 \right)  \left((1, t_\alpha) \left(\mathbf{D}^T \mathbf{V}^{-1} \mathbf{D}\right)^{-1} (1, t_\alpha)^T\right)^{1/2}
\end{equation}
by \eqref{eq-info-linear-mixed-approx} and \eqref{eq-grad-linear}, and $c_{\boldsymbol{\varsigma}}^2 = \mathbf{c}_{\boldsymbol{\varsigma}}^T \mathbf{M}_{\boldsymbol{\varsigma}}^{-1} \mathbf{c}_{\boldsymbol{\varsigma}}$ is a constant independent of $\xi$ in which $\mathbf{c}_{\boldsymbol{\varsigma}} = \partial F_{T_2}(t_\alpha) / \partial \boldsymbol{\varsigma}$ is the gradient of $F_{T_2}(t_\alpha)$ with respect to the vector $\boldsymbol{\varsigma}$ of variance parameters.

The criterion \eqref{eq-avar-gamma-linear} is a weighted sum of the optimality criteria for the single components and constitutes, hence, a compound criterion, where the weights depend on both vectors $\boldsymbol{\beta}_1$ and $\boldsymbol{\beta}_2$ of location parameters as well as on the variance parameters $\boldsymbol{\varsigma}$ of the linear mixed model component, in general. Due to convexity the optimal weight $w^*$ for the system lies in the range of the optimal weights $w_1^*$ and $w_2^*$ for the components, $\min\{w_1^*, w_2^*\} \leq w^* \leq \max\{w_1^*, w_2^*\}$.

\subsection{Numerical example}
\label{sec:2.5}
In this example we consider an accelerated degradation experiment with two independent failure components in which the first component follows a gamma process and the second is described by a linear mixed model with random intercept as described in Subsection~\ref{sec:2.1}.
The gamma process is specified as in Subsections~\ref{sec:1.4} and \ref{sec:2.5gammagamma} with scale parameter $\nu_1 = 1$, degradation threshold $z_{10} = 5.16$ and nominal values $\beta_{10} = 0.23$ for the intercept and $\beta_{11} = 0.53$ for the slope.
For the linear model we assume a degradation threshold $y_{20} = 3.73$ and nominal values $\beta_{20} = 2.35$ for the aggregate intercept, $\beta_{21} = 0.06$ for the slope in the stress variable $x$, $\beta_{22} = 0.28$ for the slope in time $t$, $\beta_{23} = 0.04$ for the stress-time interaction $xt$, $\sigma_0 = 0.08$ for the standard deviation of the random intercept, and $\sigma_\varepsilon = 0.09$ for the standard deviation of measurement errors.
As before the standardized normal use condition is $x_u = - 0.40$ and both degradation processes are measured at the $k = 4$ standardized time points $t_j = 0.25$, $0.5$, $0.75$ and $1$ with time intervals of constant length $\Delta = 0.25$.
Additionally, the degradation of the second component is measured initially at $t_0 = 0$, i.\,e., at the beginning of the experiment.
Also in the present setting we will be interested in estimating the median failure time $t_{0.5}$.

The distribution function $F_T(t)$ of the combined failure time $T$ is plotted in Figure~\ref{fig-failuretime-gammalinear} together with the distribution functions $F_{T_1}(t)$ and $F_{T_2}(t)$ of the failure times $T_1$ and $T_2$ in the components. 
The median failure time $t_{0.5} = 4.99$ satisfying $F_T(t_{0.5}) = 1/2$ is indicated there by a dashed vertical line.

For estimating the median failure time, also here the optimal design is sought numerically by means of the multiplicative algorithm on an equidistant grid of step size $0.01$ on the design region. 
As in the univariate case the algorithm indicates that the optimal design $\xi^*$ is of the form $\xi_{w}$. Under this premise the optimal value of $w^*$ can be determined by a simple line search on a sufficiently dense grid. The resulting optimal designs which assigns optimal weights $w^* = 0.78$ to the lowest stress level $x = 0$ and $1 - w^* = 0.22$ to the highest stress level $x = 1$ in the design region.

To assess the robustness of the locally optimal design we examine how the optimal weight $w^*$ varies when the underlying parameter values are modified. 
Computations indicate that the optimal weight does not change much in the nominal values of the parameters $\boldsymbol{\beta}_2$ and $\boldsymbol{\varsigma}$ for the linear mixed effects degradation model in the second component. 
This property is in accordance with the fact that the design criterion depends on the values of $\boldsymbol{\beta}_2$ and $\boldsymbol{\varsigma}$ only through the weighting factors $c_1^2$ and $c_2^2$ while the marginal information matrix $\mathbf{M}(\xi)$ does not. 
However, similar to the univariate case, there may be moderate changes with respect to the parameters $\boldsymbol{\beta}_1$ of the gamma degradation model in the first component.
Additionally, the optimal weight $w^*$ may switch between the marginal optimal weights $w_1^*$ and $w_2^*$ for the marginal failure models depending on which of the marginal failure modes is dominant in the bivaraite system.

\begin{figure}
	\centering
	\begin{minipage}[b]{0.39\textwidth}
		\includegraphics[width=\textwidth]{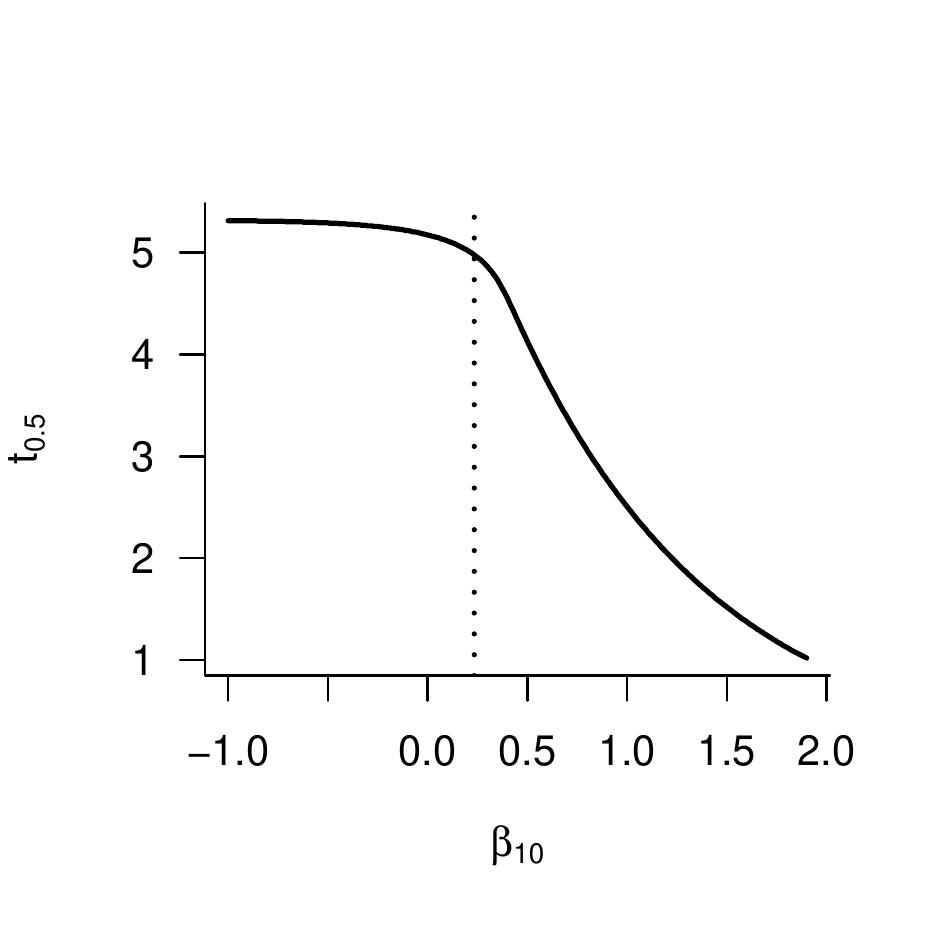}
		\caption{Dependence of $t_{0.5}$ on $\beta_{10}$ for the example in Subsection~\ref{sec:2.5}}
		\label{fig_failure-quantile_vs_beta10_bivariate}
	\end{minipage}
	\hfill
	\begin{minipage}[b]{0.4\textwidth}
		\includegraphics[width=\textwidth]{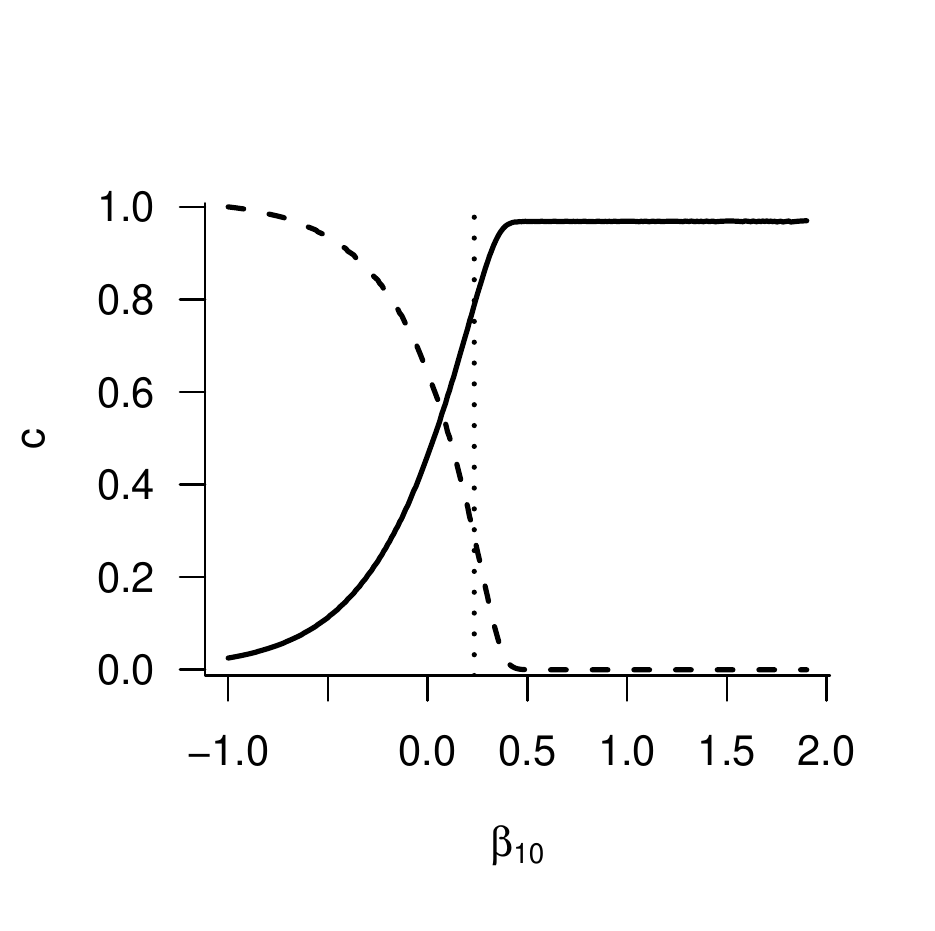}
		\caption{Dependence of the coefficients $c_1$ (solid line) and $c_2$ (dashed line, standardized) on $\beta_{10}$ for the example in Subsection~\ref{sec:2.5}}
		\label{fig_coeffs_vs_beta10_bivariate}
	\end{minipage}
\end{figure}

We will demonstrate this behavior in the case when the intercept $\beta_{10}$ of the gamma model component varies while all other parameters are fixed to their nominal values.
In Figure~\ref{fig_failure-quantile_vs_beta10_bivariate} we plot the median failure time $t_{0.5}$ and the weighting coefficients $c_1$ and $c_2$, respectively, in dependence on $\beta_{10}$.
For negative values of $\beta_{10}$, the failure time $T_2$ of the first component decreases and the failure of the bivariate system is dominated by the second component.
Then the median failure time $t_{0.5}$ approaches its marginal counterpart $5.32$ in the second  component. 
For increasing values of $\beta_{10}$, the failure of the first component becomes dominant and the median failure time $t_{0.5}$ behaves as in the marginal model for the first component.
In particular, $t_{0.5}$ is decreasing in $\beta_{10}$ and becomes smaller than $1$ for $\beta_1 > 1.92$.
Hence, only values $\beta_{10} \leq 1.92$ are reasonable to be considered because otherwise no acceleration would be required to obtain failure due to degradation under normal use conditions.
The dominance of the failure components is also reflected in Figure~\ref{fig_coeffs_vs_beta10_bivariate} where the weighting coefficients $c_1$ and $c_2$ are shown in dependence on $\beta_{10}$.
There the second coefficient $c_2$ is standardized by its maximum for purposes of comparison.
For negative values of $\beta_{10}$, the coefficient $c_2$ of the second component dominates the asymptotic variance \eqref{eq-avar-gamma-linear} while the dominance is is reversed for $\beta_{10} > 0.5$.

\begin{figure}
	\centering
	\begin{minipage}[b]{0.42\textwidth}
		\includegraphics[width=\textwidth]{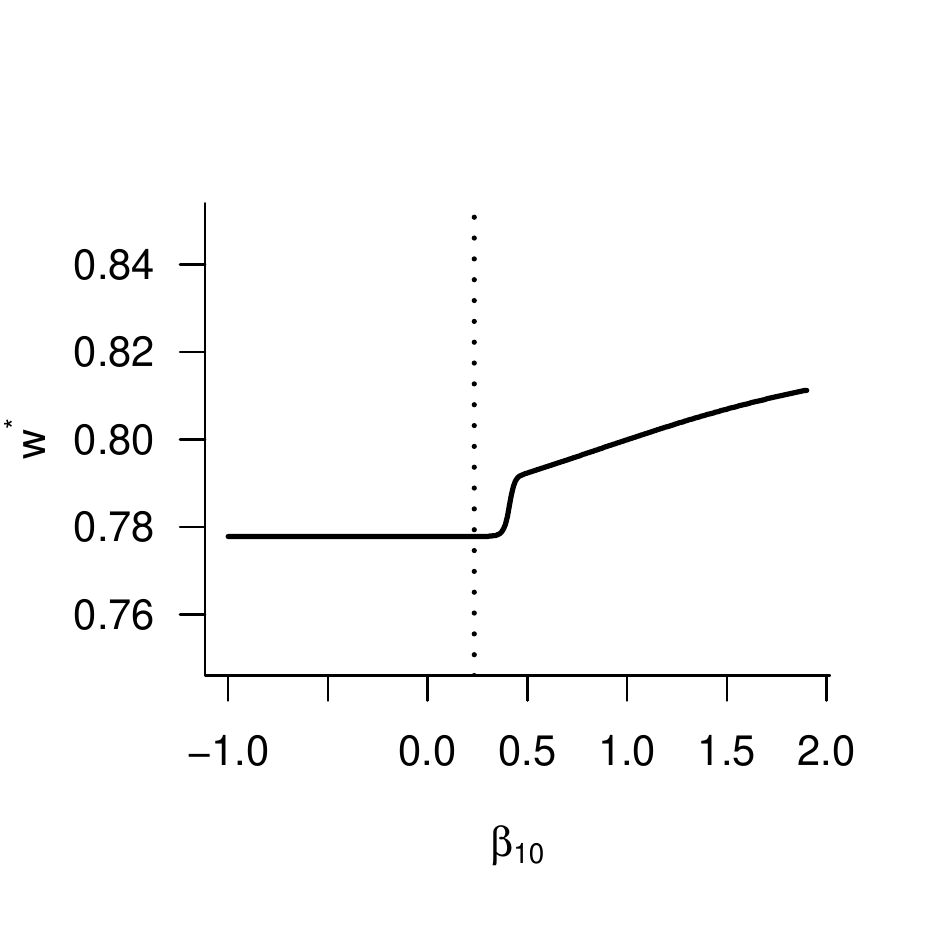}
		\caption{Dependence of $w^*$ on $\beta_{10}$ for the example in Subsection~\ref{sec:2.5}}
		\label{fig_weight_vs_beta10_bivariate}
	\end{minipage}
	\hfill
	\begin{minipage}[b]{0.425\textwidth}
		\includegraphics[width=\textwidth]{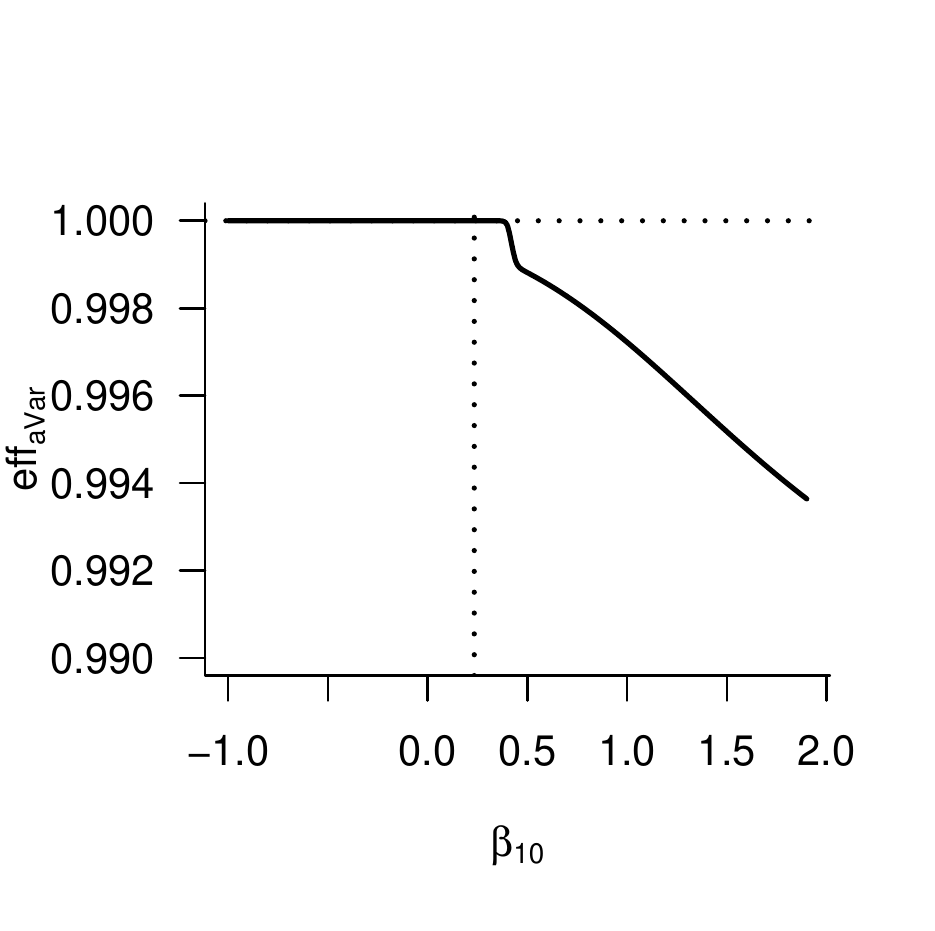}
		\caption{Efficiency of $\xi^*$ in dependence on $\beta_{10}$ for the example in Subsection~\ref{sec:2.5}}
		\label{fig_eff_vs_beta10_bivariate}
	\end{minipage}
\end{figure}

This change in dominance has also an impact on the optimal weights $w^*$ as exhibited in Figure~\ref{fig_weight_vs_beta10_bivariate}.
For negative values of $\beta_{10}$, the optimal weight $w^*$ coincides with its marginal counterpart in the second component while, for $\beta_{10} > 0.5$, the optimal weight $w^*$ is as in the univariate model for the first component (see Subsection~\ref{sec:1.4}).
Caused by the change in dominance, there is a small, but pronounced change in the optimal weight when $\beta_{10}$ varies from $0.35$ to $0.50$.
This shift is also visible in the efficiency of the locally optimal design $\xi^*$ at the given nominal values when the intercept parameter $\beta_{10}$ is misspecified, as shown in Figure~\ref{fig_eff_vs_beta10_bivariate}.
For values of $\beta_{10}$ less than the nominal value $\beta_{10} = 0.23$, the locally optimal design $\xi^*$ has an efficiency of nearly $1$, up to $\beta_{10} = 0.35$. 
Then there is a small, instantaneous decrease in efficiency to $0.999$ for $\beta_{10}$ between $0.35$ and $0.50$.
For larger values of $\beta_{10}$ the efficiency smoothly decreases as in the univariate model for the first component (see Subsection~\ref{sec:1.4}).
For the maximal value $\beta_{10} = 1.92$ the efficiency of $\xi^*$ is still remarkably high with a value of about $0.9936$.
In all of Figures~\ref{fig_failure-quantile_vs_beta10_bivariate} to \ref{fig_eff_vs_beta10_bivariate} the nominal value $\beta_{10} = 0.23$ is indicated by a dotted vertical line.

With respect to the slope parameter $\beta_{20} \leq 0$ of the gamma component, the failure is dominated by the second component.
Hence, neither the optimal weight is affected by $\beta_{20}$, nor the efficiency of the locally optimal design $\xi^*$ differs reasonably from $1$.
In total, the locally optimal design $\xi^*$ at the given nominal values appears to be robust against misspecifications of the parameters within a meaningful range.

\section{Concluding remarks}
\label{sec1.4444}
The design stage of highly reliable systems requires a sophisticated assessment of the reliability related properties of the product. 
One approach to handle this issue is to conduct accelerated degradation testing. 
Accelerated degradation tests have the advantage to provide an estimation of lifetime and reliability of the system under study in a relatively short testing time. 
The majority of existing literature deals with this issue by considering a single failure mode, which may not be sufficiently representative in many cases. 

In this work, we propose optimal experimental designs for ADTs with a single response components and extend it to the case of multiple response components with repeated measures. Two bivariate degradation models are considered. The marginal degradation functions are described by two gamma process models in the first bivariate model, and a gamma process with a linear model with a random intercept in the second one.
In this context it is desirable to estimate certain quantiles of the joint failure time distribution as a characteristic of the reliability of the product.
The purpose of optimal experimental design is to find the best settings for the stress variable to obtain most accurate estimates for these quantities.

In the present model for accelerated degradation testing, it is assumed that stress remains constant within each testing unit during the whole period of experimental measurements but may vary between units.
Hence, in the corresponding experiment a cross-sectional design between units has to be chosen for the stress variable(s) while for repeated measurements the time variable varies according to a longitudinal time plan within units. In particular, the same time plan for measurements is used for all units in the test.
It is further assumed that the marginal response components are uncorrelated.  

The multiplicative algorithm is utilized to obtain optimal experimental designs for the single response case as well as the two bivariate degradation models. The sensitivity analysis shows that the optimal designs of the univariate model as well as the bivariate model with two marginal gamma processes are robust against misspecifications of the corresponding parameter vectors and depend mainly on the normal use condition of the stress variable. For the bivariate model with two different marginal models the sensitivity analysis establishes that the resulting optimal design is slightly dependent on the nominal parameter values.

 Although only gamma processes and LMEM are considered as marginal degradation models here, the underlying methods can be extended to other marginal failure modes, like 
Wiener processes, inverse Gaussian processes and non-linear mixed effects degradation models. 
Another object of interest would be to consider optimality criteria accounting for simultaneous estimation of various characteristics of the failure time distribution.

\section*{Acknowledgment}
The work of the first author has been supported by a research grant of the German Academic Exchange Service (DAAD) under the title (Promotionen in Deutschland, (2017-18)/ID-57299294)). The authors are indebted to Professor Norbert Gaffke of the Institute of Mathematical Stochastic at the University of Magdeburg for his valuable comments and constructive suggestions during the work on this paper.

\newpage
\section*{References}
\bibliographystyle{apa}
\bibliography{Reference_GammaLinear}

\end{document}